**Percolation lithography: Tuning and freezing disorder in 3D photonic crystals using partial wetting and drying**

Ian B. Burgess[1,2*], Navid Abedzadeh[3], Theresa M. Kay[2,3], Anna V. Shneidman[4], Derek J. Cranshaw[3], Marko Lončar[3*], Joanna Aizenberg[2,3,4,5*]

[1]Leslie Dan Faculty of Pharmacy, University of Toronto, Toronto, Ontario Canada
[2]Wyss Institute for Biologically Inspired Engineering, Harvard University, Cambridge, MA, USA
[3]School of Engineering and Applied Sciences, Harvard University, Cambridge, MA, USA
[4]Department of Chemistry and Chemical Biology, Harvard University, Cambridge, MA, USA
[5]Kavli Institute for Bionano Sciences and Technology, Harvard University, Cambridge, MA, USA
*email: ib.burgess@utoronto.ca, loncar@seas.harvard.edu, jaiz@seas.harvard.edu

**Although complex, hierarchical nanoscale geometries with tailored degrees of disorder are commonly found in biological systems, few simple self-assembly routes to fabricating synthetic analogues have been identified. We present two techniques that exploit basic capillary phenomena to finely control disorder in porous 3D photonic crystals, leading to complex and hierarchical geometries. In the first, we exposed the structures to mixtures of ethanol and water that partially wet their pores, where small adjustments to the ethanol content controlled the degree of partial wetting. In the second, we infiltrated the structures with thin films of volatile alkanes and observed a sequence of partial infiltration patterns as the liquid evaporated. In both cases, macroscopic symmetry breaking was driven by subtle sub-wavelength variations in the pore geometry that directed site-selective infiltration of liquids. The resulting patterns, well described by percolation theory, had significant effects on the photonic structures' optical properties, including the wavelength-dependence and angular dependence of scattering. Incorporating cross-linkable resins into our liquids, we were able create permanent photonic structures with these properties by freezing in place the filling patterns at arbitrary degrees of partial wetting and intermediate stages of drying. These techniques illustrate the versatility of interfacial phenomena in directing and tuning self-assembly of aperiodic structures.**

Several decades of studying periodic photonic systems, from thin-films to 3D photonic crystals, have produced numerous significant technological advances *(1-4)*. The broad experimental study of 3D photonic crystals was made possible by advances in nanoscale manufacturing *(5-11)*, including the large-scale production periodic photonic structures by self-assembly methods *(8-11)*. More recently the study of aperiodic photonic structures, ranging from small defects in periodic lattices *(12-14)* to disordered and quasi-periodic structures *(15-19)*, has led to the prediction and discovery of a much more diverse range of optical effects, with wide-ranging applications from random lasers *(20)* to structurally colored materials with precisely controllable wavelength and angular dependence of scattering *(21)*. While in 1D and 2D, arbitrary structures with designer symmetries and optical properties can be fabricated by planar nanofabrication techniques *(22-24)*, our ability to fabricate precisely controlled aperiodic, disordered or partially disordered 3D photonic structures remains limited.

In contrast, photonic structures in biological organisms, including butterflies, beetles and birds, commonly exploit complex 3D periodic and aperiodic morphologies with varying degrees of disorder to produce complex optical properties *(25-29)*. These structures are all produced by self-assembly processes, often exploiting simple interfacial physics under controlled conditions to make and break symmetries in tailored ways *(30)*. Inspired by these biological examples, we show here how simple capillary phenomena driven by interfacial physics, the partial imbibition of a liquid into or drainage of a liquid from a porous 3D photonic crystal, can be used to create complex and hierarchical 3D photonic structures with a controllable degree of disorder. Liquids create this tunable disorder in the structures under conditions where subtle variations in the pore morphology from one lattice site to the next (on scales much smaller than visible wavelengths) lead to selective infiltration or drainage *(31,32)*. These modifications have significant effects on the wavelength-dependence and angular dependence of scattering. Incorporating cross-linkable resins into our liquids, we were able to freeze into place these structures at arbitrary stages of incomplete wetting or drying.

The 3D photonic crystals used in this work were inverse-opal films (IOFs) made of silica, and were fabricated on silicon substrates using an evaporative co-assembly method described previously *(33)*. IOFs produced by this technique contain regularly spaced spherical voids ($d \sim 300$ nm) arranged in a face-centered-cubic lattice with a single crystallographic orientation and very low defect density across centimeter scales. Neighbouring voids are connected by small circular openings (necks). These necks have a highly re-entrant geometry that encourages contact-line pinning as a liquid menisci pass through them *(31,34)*. At the necks, there is a significant free-energy barrier for the contact line to overcome during imbibition and drainage *(32)*. While in an ideal photonic crystal, the pore geometry is exactly the same at all lattice sites, small variation in the pore and neck geometry is observable in our IOFs *(32)*. Variation in the neck geometry translates into subtle differences in the strength of these free energy barriers between different lattice sites. The scale of this broken symmetry is much smaller than visible wavelengths and therefore it has little direct effect on the IOFs optical properties. However, under the right conditions these variations selectively direct the imbibition or drainage of liquids, superimposing macroscopic hierarchical structures onto the underlying periodic photonic crystal.

When an IOF is immersed in liquid, the meniscus will be able to move from one pore to the next if the intrinsic contact angle ($\theta_c$) is smaller than the re-entrant neck angle ($\phi_0$) *(31,32,34)*. The IOFs contain pores whose neck angles vary randomly according to a fairly narrow distribution (e.g. $\pm 3°$ *(32)*). Most liquids are likely to have $\theta_c$ far outside the distribution of neck angles, leading to penetration that is either complete or non-existent *(32)*. However, when IOFs are immersed in liquids whose $\theta_c$ falls within the narrow range defined by the neck angle distribution, incomplete wetting occurs, as some necks will pin the meniscus while others will not. At equilibrium, the liquid will occupy the network of all paths from the outer surface that are connected by necks that do not pin ($\phi_0 > \theta_c$). Simulations on a 2D IOF in Fig. 1A illustrate what these defect patterns look like. The percolation simulation used the model described in Ref. *(32)*, where the structure was assigned a random distribution of neck angles with a mean of $19.6°$ and a standard deviation of $3.2°$, values that describe typical IOFs *(32)*. Since the refractive-index contrast is diminished in the liquid-filled pores, filling a pore effectively removes it from the lattice structure. Thus, the effect of partial wetting is to impose patterns of missing lattice sites onto the photonic structure. These patterns of disorder, described by percolation theory *(32)*, can be tuned by adjusting $\theta_c$ for the liquid.

Similar defect patterns can be superimposed on the photonic crystal structure during evaporation of a liquid that has completely filled the porous network. In contrast to the static equilibrium defect patterns created via partial wetting, defect patterns observed during drying evolve continuously over time as liquid evaporates, with the full spectrum of patterns between complete filling and completely empty pores occurring during the course of each experiment. These patterns are well understood via the theory of invasion percolation *(35,36)*. Similar to the case of partial wetting, random variations in the neck geometry produce a pinning effect on the receding meniscus during drainage whose strength varies across different necks. This underlying sub-wavelength broken symmetry therefore directs the recession of the meniscus into a series of random walks through the structure, creating an evolving sequence of hierarchical partial filling patterns. Fig. 1B illustrates the nature of this partial filling sequence, based on a simulation of liquid drying from a 2D IOF (see Methods for simulation details).

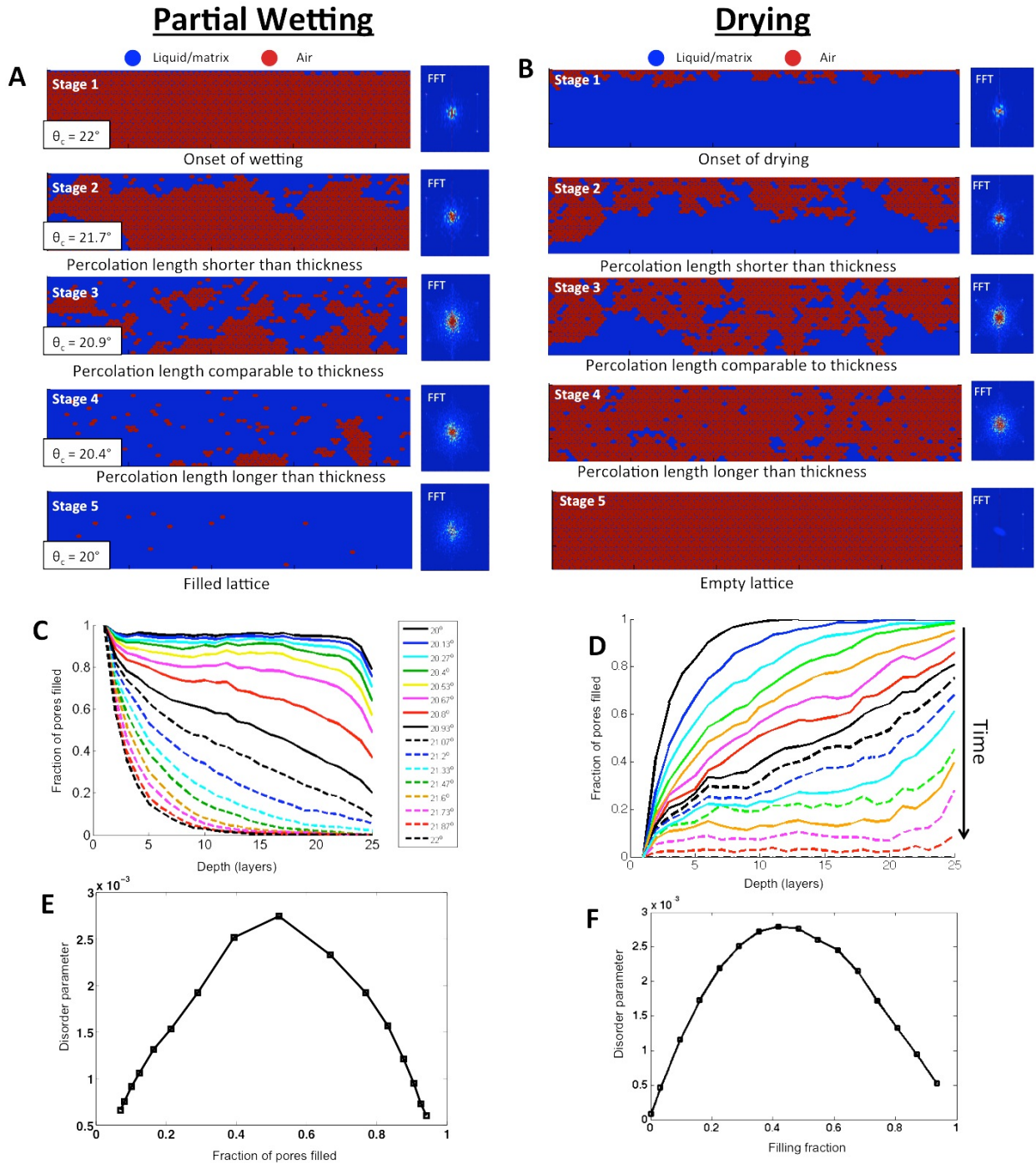

**Figure 1 – Simulated evolution of disorder in inverse-opal films (IOFs) due to partial wetting and drying.** A,B: Percolation simulations of 2D IOFs showing the progression of defect patterns occurring during partial wetting as a function of the liquid's intrinsic contact angle (A) and drying as it evolves in time (B). Fourier transforms (FTs) of the refractive-index map are shown to the right of each image. The density of the FT outside of the vertical axis expands and contracts with the degree of disorder. C,D: Fraction of filled pores at each layer evolving as a function of $\theta_c$ during partial wetting (C) and as a time progression during drying (D). E,F: Disorder, quantified from the lattice FTs as the fraction of spectral density in the first Brillouin zone outside of the vertical axis, plotted as a function of the overall filling fraction, for partial wetting (E) and drying (F). Maximum disorder occurs in both cases when half of the pores are filled.

Simulated refractive-index maps (left) and their Fourier Transforms (FTs, right) derived from our simulations of partial wetting and drying on 2D IOFs at five characteristics stages are shown in Fig. 1A (partial wetting) and Fig. 1B (drying). Figure 1C,D shows the evolution of the fraction of pores filled as a function of depth for partial wetting (Fig. 1C) and drying (Fig. 1D). The evolution of partial filling patterns in partial wetting (Fig. 1A) and drying (Fig. 1B) are very similar in nature, except that the locations of liquid and air are interchanged. In both cases, pattern progressions can be divided into five distinct stages, which are distinguished by the relationship between the percolation length and the film thickness. Stage 1 marks the onset of pore filling (partial wetting) or pore emptying (drying) where the percolating front has just begun to invade the initially ordered lattice. At Stage 2, the wetting/drying front penetrates significantly into the lattice, but the percolation length is shorter than the total thickness of the film and the deepest layers remain relatively unperturbed. When the percolation length is shorter than or comparable to the film thickness (Stages 1-3), the fraction of pores filled by the invading species (partial wetting: liquid-filled pores; drying: air-filled pores) decreases with depth, reflecting the decreasing probability of connected paths extending to pores at larger depths (see Fig. 1C – lower dashed curves, Fig. 1D – upper solid curves). We define Stage 3 as the point where the percolation length is comparable to the thickness, the lattice is half-filled and the invading species (partial wetting: liquid-filled pores; drying: air-filled pores) becomes the majority in both wetting and drying. At Stage 4, the percolation length is larger than the thickness and the receding species (partial wetting: air-filled pores; drying: liquid-filled pores) become increasingly isolated defects that are homogeneously distributed in the film. Once the percolation length has grown well past the thickness of the film (eventually crossing the percolation threshold where the percolation length, $L \rightarrow \infty$) the probability of finding connected paths to the surface no longer varies with depth and the filling fraction becomes homogeneous (see Fig. 1C – upper solid curves, Fig 1D – lower dashed curves). Stage 5 marks the completion of wetting or drying and recovery of a completely liquid-filled lattice (partial wetting) or defect-free photonic crystal (drying), both of which are ordered structures.

In both partial wetting and drying, the degree of disorder initially increases as the percolation progresses from an ordered state at the outset (wetting: empty photonic crystal; drying: uniformly filled lattice). In both scenarios, the degree of disorder reaches a maximum at the point of half filling (Stage 3, Fig. 1E,F). Once the percolation progresses past this point disorder decreases, as the invading species become the majority (partial wetting: liquid-filled pores; drying: air-filled pores) and the receding species become the defect centers (partial wetting: air-filled pores; drying: liquid-filled pores). Figure 1E,F plots a measure of disorder in the lattice as a function of the overall filling percentage, showing its maximization at 50% filling in both scenarios. We quantified the degree of disorder in our simulations using fraction of the density in the refractive index map's FT that lies outside the vertical axis in the first Brillouin Zone. This region is chosen to separate the contributions of lattice disorder from the FT density associated with reciprocal lattice sites, the zero-spatial-frequency component at the origin, and the contributions along the vertical axis coming from the top and bottom interfaces of the film.

We experimentally tuned the degree of partial wetting by immersing IOFs, uniformly functionalized with n-decyl trichlorosilane (DEC) to render their pores hydrophobic, in different mixtures of water and ethanol (Fig. 2A-C). Adjusting the ratio of water and ethanol in the mixtures allowed us to create a continuum of $\theta_c$ values in the pores *(31,32)*. We experimentally observed the time-evolution of disorder during drying after filling IOFs with thin films of alkane liquids (decane, undecane, dodecane). To observe the desired effects, we required initial conditions where the drying front would hit the top surface of the IOF over a large area at the same time. This meant that the pores had to be initially completely filled with a stable film of liquid (zero contact angle) and minimal excess liquid on top of the film. To create these conditions, DEC-functionalized IOFs were immersed in alkane liquids and then immediately flushed under running water. The water, an immiscible liquid having less affinity for the DEC-coated pore surfaces than the alkanes, trapped the alkane inside the pores while removing the excess alkane on top. After flushing, all excess water slid easily off the surface as the retained alkane acted as a lubricant *(37,38)*, leaving only the IOF with alkane inside the pores (and a very thin liquid film above). This enabled drying to occur concurrently across the entire film, and allowed us to visualize percolation-induced disorder on a large scale (Fig. 2D-F). Specific alkanes were chosen based on their volatility, allowing us to set the ideal drying timescale for the different experiments described below.

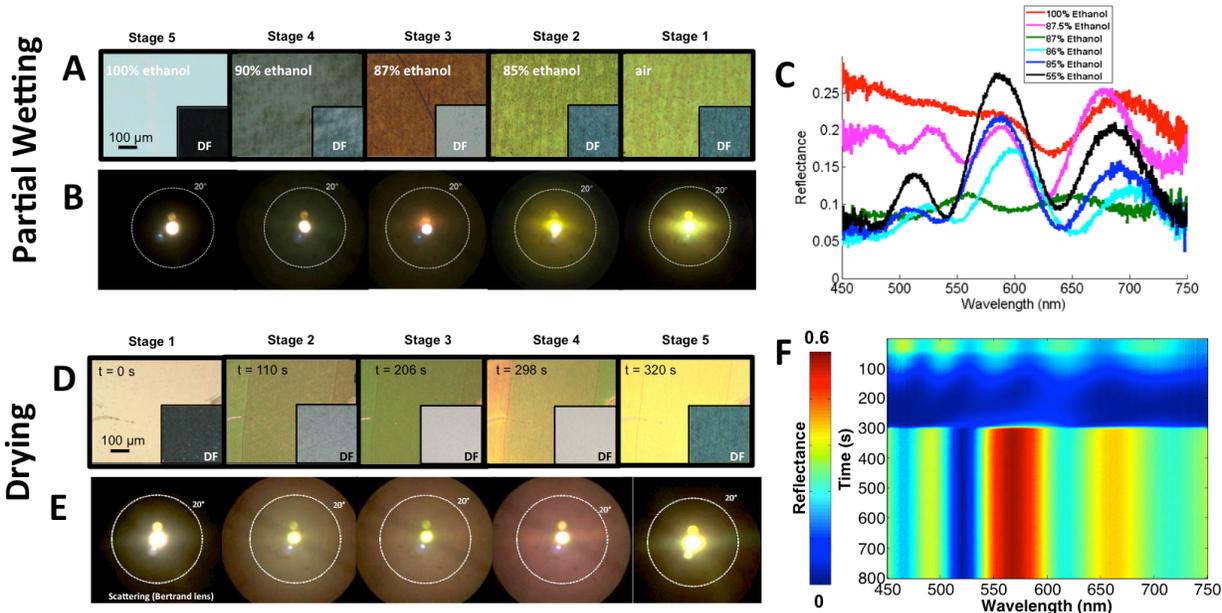

**Figure 2 – Experimental evolution of color and scattering during partial wetting and drying.** (A) Bright-field and dark field (insets) images of a DEC-functionalized IOF (9 layers) immersed in different ethanol-water mixtures. (B) Corresponding Bertrand lens images showing angular distribution of scattering. (C) Reflectance spectra at normal incidence of an IOF (9 layers) immersed in various ethanol-water mixtures, showing the effects of partial wetting. (D) Time lapse bright field and dark field (insets) images of an IOF (9-layers) as dodecane evaporates from the pores. (E) Corresponding Bertrand lens images showing angular distribution of scattering. (F) Time evolution of reflectance at normal incidence for an IOF (9 layers).

The experimental evolution of IOFs' optical properties after partial wetting (immersion in different mixtures of water and ethanol) and during drying (the time-evolution of dodecane evaporation) are shown in Fig. 2 (A-C: partial wetting, D-F: drying) for IOFs having 9 close-packed layers in thickness. Figure 2A,D show bright-field and dark-field images (Fig. 2A: wetting, Fig. 2D: drying) of the IOFs at ethanol concentrations (wetting) and time points (drying) corresponding to the five stages shown in Fig. 1A,B and discussed above. These images are accompanied by images taken through a Bertrand lens *(39)* that map the angular distribution of scattering at each stage (Fig. 2B: wetting, Fig. 2E: drying). Supplementary Movies M1-M3 show the complete time evolution of the dodecane drying experiments in bright-field (M1), dark-field (M2) and imaged through the Bertrand lens (M3). Reflection spectra (normal incidence) for partial wetting are shown for several ethanol-water mixtures that cover the five stages (Fig. 2C). The time-resolved reflection spectrum for dodecane drying is shown in Fig. 2F.

As partial filling patterns produced in wetting and drying (compare Figs. 1A and 1B) are very similar in nature, there are also many commonalities in the optical effects they produce (compare Figs. 2A-C and 2D-F). In both scenarios, the spectral evolution has of two main characteristics that occur simultaneously. First, the degree of diffuse scattering increases and then decreases after reaching a maximum (Stage 3), following the expected evolution in the degree of disorder (Fig. 1E,F) as the percolation processes evolve. Increased diffuse scattering results in a decrease in brightness of bright-field images (Fig. 2A,D), an increase in brightness of dark-field images (Fig. 2A,D, insets), a broadening of the bright area in the Bertrand-lens images (Fig. 2B,E), and a decrease in the total reflectance at normal incidence evident in the spectra (Fig. 2C,F). Second, the qualitative characteristics of the spectra (and associated color) transition from having a Bragg resonance when the pores are predominantly air-filled (wetting: Stage 1; drying: Stage 5) to having thin-film characteristics (wetting: Stage 5; drying: Stage 1), when the pores are predominantly liquid-filled, reflecting the relatively close refractive-index match between the liquid and the matrix.

The completely filled thin-film stage reveals the most important difference between the spectral response in partial wetting and drying experiments, which arises from the region directly above the structure being liquid at all stages in our wetting experiments and being air at all stages in drying experiments. This prominent upper interface in the drying experiments leads to enhanced appearance of thin-film fringes in the completely wet (beginning) stage of drying experiments (Fig. 2F) compared to the completely wet IOFs in wetting experiments (Fig. 2C). The presence of a greater refractive-index contrast at the top surface in drying also leads to the stronger overall reflectance compared to partial wetting at all stages.

The thin-film character of the reflectance spectrum is maintained for much of the majority-liquid-filled stage (wetting: Stage 4; drying: Stage 2) between complete liquid filling (wetting: Stage 5; drying: Stage 1) and the point of maximum scattering (Stage 3). Likewise, a prominent peak corresponding to the Bragg resonance is also observed for much of the majority-air-filled stage (wetting: Stage 2; drying: Stage 4) between maximum scattering (Stage 3) and complete air filling (wetting: Stage 1; drying: Stage 5). However, side-fringes are significantly suppressed with increased liquid filling during this stage, a consequence of transverse disorder. In both transition stages (Stages 2,4) for wetting and drying experiments, redshifting of spectral features occurs with increased liquid filling (see Fig. 2C,F). This trend reflects the increase in the average optical path length between pores as more are filled with the higher-index liquid. To confirm this explanation for the observed shifts, we conducted 1D optical reflectance simulations (transfer matrix method) on the refractive index profiles generated by our percolation simulation after averaging the refractive index across all transverse co-ordinates at each depth. This simulation (Fig. S1, supplementary information), which serves to remove the effects of transverse disorder, reproduces much of the qualitative behaviour of the experimentally-observed spectral evolution, including the transition from thin film to Bragg resonance and the redshifting of spectral features with increased filling. However, it does not capture the sharp decrease in the overall reflectance intensity at intermediate filling stages, which are a result of transverse disorder.

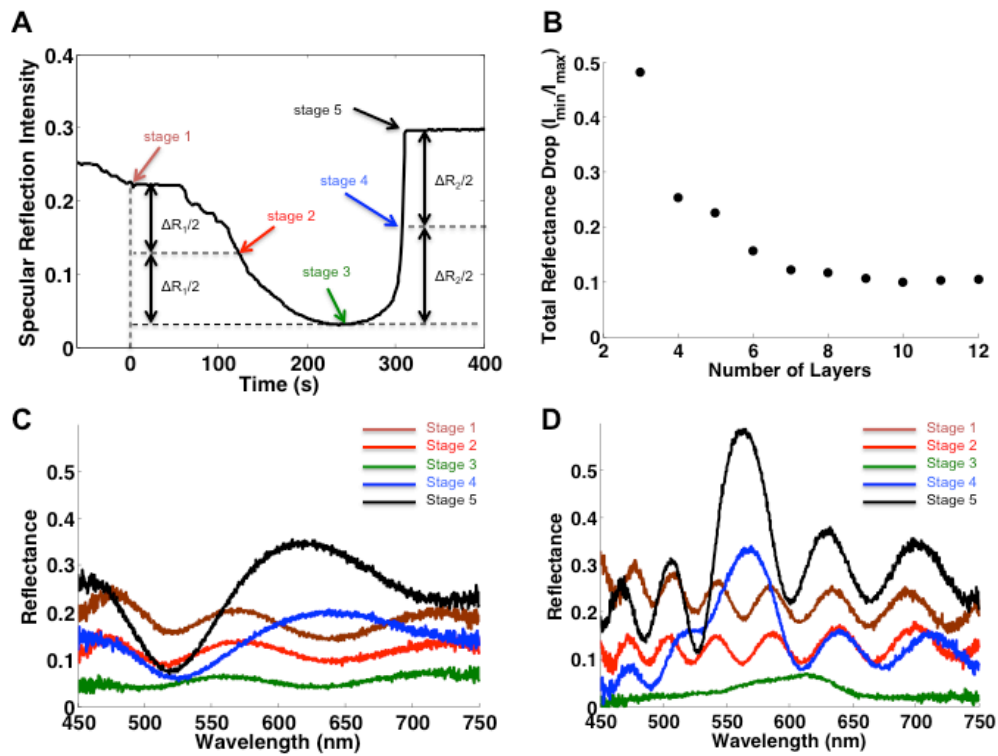

**Figure 3 – Thickness dependence of spectral evolution during drying.** (A) Time evolution of total reflectance between 450 nm and 750 nm ($t = 0$ denotes the disappearance of excess liquid above the IOF) during drying of dodecane from a 12-layer IOF. Spectral signatures were compared at five different stages defined by the total reflectance: Stage 1: disappearance of the over-layer and start of pore-emptying; Stage 2: Total reflectance decreased to halfway between Stage 1 and the minimum reflectance; Stage 3: Point of minimum reflectance; Stage 4: Total reflectance recovered halfway from the minimum value toward the end (dry) value; Stage 5: Drying completed. (B) Thickness dependence of the total reflectance change between Stage 3 and 5. (C,D) Reflectance spectra at all five stages for a 5-layer (C) and 12-layer (D) IOF.

Changing the total thickness of the IOF had two notable consequences on the way its optical properties evolve during percolation. First, we found that the strength of transverse scattering and the suppression of reflectance when disorder was at its maximum (Stage 3) increased as film thickness increased. This trend is shown in Fig. 3A,B. Fig. 3A plots the time evolution of total reflectance across the visible spectrum (450-750 nm) during a dodecane drying experiment on a 12-layer IOF. Fig. 3B plots the ratio of the minimum reflectance ($I_{min}$, Stage 3) and maximum reflectance ($I_{max}$, Stage 5) during a drying experiment as a function of film thickness. The extent of the reflectance drop initially increases with thickness, eventually saturating at a ratio of around 1/10 for IOFs thicker than 10 layers.

Second, we found that the qualitative transition of the reflectance spectrum from thin-film character to photonic-crystal character (the appearance of a Bragg resonance) occurred after the point of minimum reflectance (maximum disorder) for the thinnest samples ($h$ < ~6 layers), while it occurred before this point for thicker samples (see Figs. S2-S10 for time-resolved reflection spectra of dodecane evaporation from IOFs having thicknesses from 3 to 11 layers). This trend is shown in Fig. 3C,D, which compares the reflectance spectrum for IOFs of 5 layers and 12 layers at the five different time points of dodecane drying: the disappearance of the over-layer and onset of percolation of the drying front into the pores (Stage 1); the reduction of total reflectance to halfway between Stage 1 and the minimum reflectance (Stage 2); the point of minimum total reflectance (Stage 3); the recovery of total reflectance to halfway between the minimum reflectance and a dry film (Stage 4); and completion of drying (Stage 5).

For the IOF containing 5 layers, the spectrum at Stage 3 looks qualitatively the same as for Stages 1 and 2, whereas for 12 layers, a Bragg resonance is prominent by Stage 3. While we expect that back-reflection should be minimized at the point of maximum disorder, when half of the pores have emptied (see Fig. 1E,F), we have no reason to expect the qualitative spectral transition to occur at this point. Instead, we hypothesize that the qualitative change in the spectrum shape and the appearance of a Bragg resonance should depend on when the typical cluster size of empty pores reaches a certain fixed number of periods ($N$) determined by the refractive-index contrast. Therefore, in thin films ($h$ < $N$), the point of minimum reflectance would occur before the appearance of a Bragg resonance, whereas the sequence would be reversed for thicker films ($h$ > $N$). This trend is in fact observed in our experiments (See Figs. 3C,D, S2-S10), from which we can estimate that this critical thickness, $N$, is about 6 layers. For thinner IOFs (Figs. 3C, S2-S4), the spectral shape at Stage 3 has the same appearance as earlier stages (Stages 1 and 2) consistent with thin-film behaviour. For these thicknesses the qualitative change in the peak shape occur after the point of minimum reflectance (Stage 3). For thicker films (Figs. 3D, S6-S10), prominent single peaks characteristic of a Bragg resonance are observed at the point of minimum reflectance (Stage 3) with the significant qualitative changes in the peak shape occurring earlier.

Although the time scale associated with the drying process increased with the liquid's volatility, the sequence of partially disordered defect patterns and their spectral signatures did not depend on volatility and could be well reproduced from one run to the next and across liquids with very different volatility. Fig. S11 shows a comparison of the spectral signatures of decane, undecane and dodecane drying from an IOF taken at the five Stages whose time-points were specified using total reflectance (see Fig. 3A). While the time taken to reach each stage varied greatly between the liquids, the spectral signatures themselves showed relatively little difference. Assuming the initial condition that the drying front invades from the top of the film, enforced by the manner in which we removed the liquid over-layer with water purging, the order in which the pores empty as an IOF dries should be driven predominantly by subtle asymmetries in the pore and neck structure as opposed to depending strongly on properties of the liquid (e.g. volatility, surface tension) *(36)*.

## Freezing intermediate stages of drying

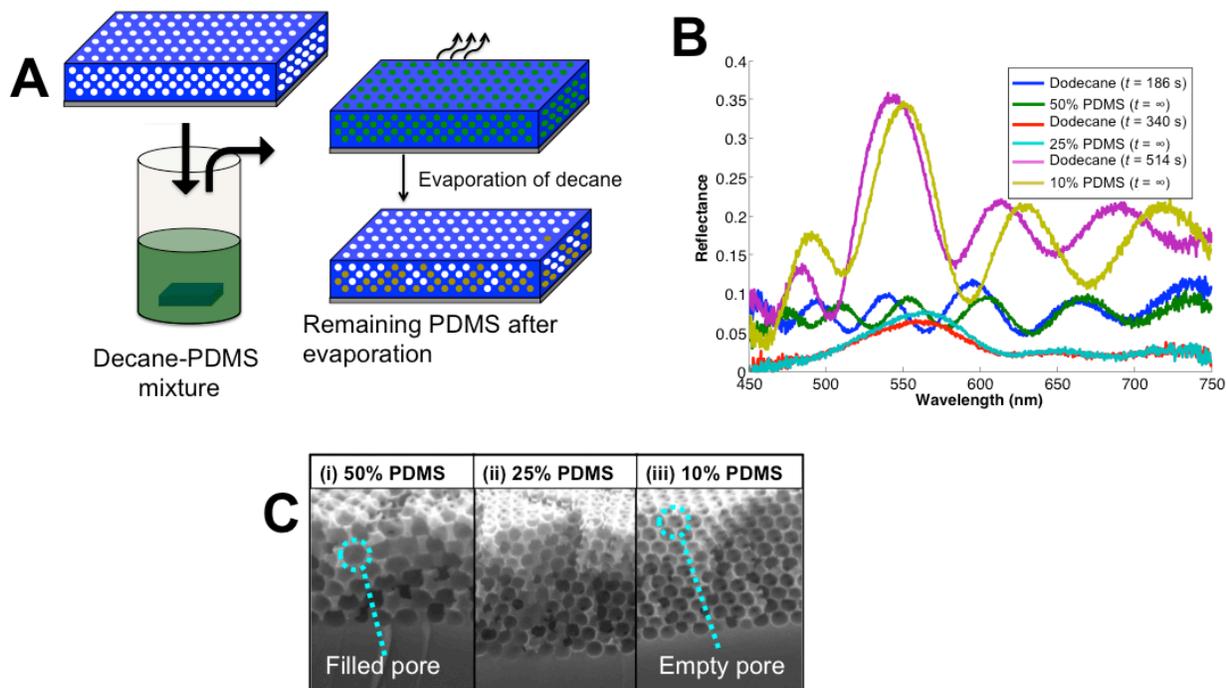

**Figure 4 – Freezing pore-filling profile during arbitrary stages of drying.** (A) Schematic showing the procedure. An IOF is wetted in a mixture of PDMS resin (Sylgard 184) and decane. After excess liquid is removed, evaporation-driven percolation proceeds until only the non-volatile PDMS remains. The resin is then cured to make a permanent structure. The initial decane:PDMS volume fraction determines the stage of drying in which the structure is frozen. (B) Reflection spectra of IOFs (9 layers) at various time-points as 100% dodecane evaporates from the structure, compared to frozen partially PDMS-filled structures made from decane-PDMS mixtures at concentrations (i, $t$ = 186 s, ii, $t$ = 340 s, iii, $t$ = 514 s) designed to replicate the corresponding stages of dodecane drying (i, 50% PDMS, ii, 25% PDMS, iii, 10% PDMS). (C) SEM images of cross-sections of the partially PDMS-filled IOFs from (B).

Using this invariance principle we engineered the volatility profile of our liquids to permanently freeze in place the defect profile at arbitrary intermediate stages of the drying process. We filled IOFs with mixtures of decane and polydimethylsiloxane (PDMS) resin (Sylgard 184, see schematic in Fig. 4A). After allowing drying to proceed until the decane component evaporated, we thermally cured the PDMS to solidify the structure in place. The volume fraction of non-volatile PDMS resin determined the stage of percolation in which the structure would freeze and thus its optical properties. Fig. 4B shows the reproduction of the optical signatures of IOFs during different time-points of dodecane drying (approximating Stages 2, 3 and 4) replicated in permanent partially disordered IOFs created from different starting PDMS concentrations. Scanning electron micrographs of cross-sections of these IOFs, showing the permanent partially filling structures are shown in Fig. 4C. The initial volume fractions of PDMS were significantly lower than the final fraction of pores filled (e.g. 25% PDMS produced roughly 50% filling), as is evident from the images in Fig. 4C. This disparity reflects the slight excess of liquid present as a thin film that remains on top of the structure at the start of evaporation after our water-flushing step *(37,38)*.

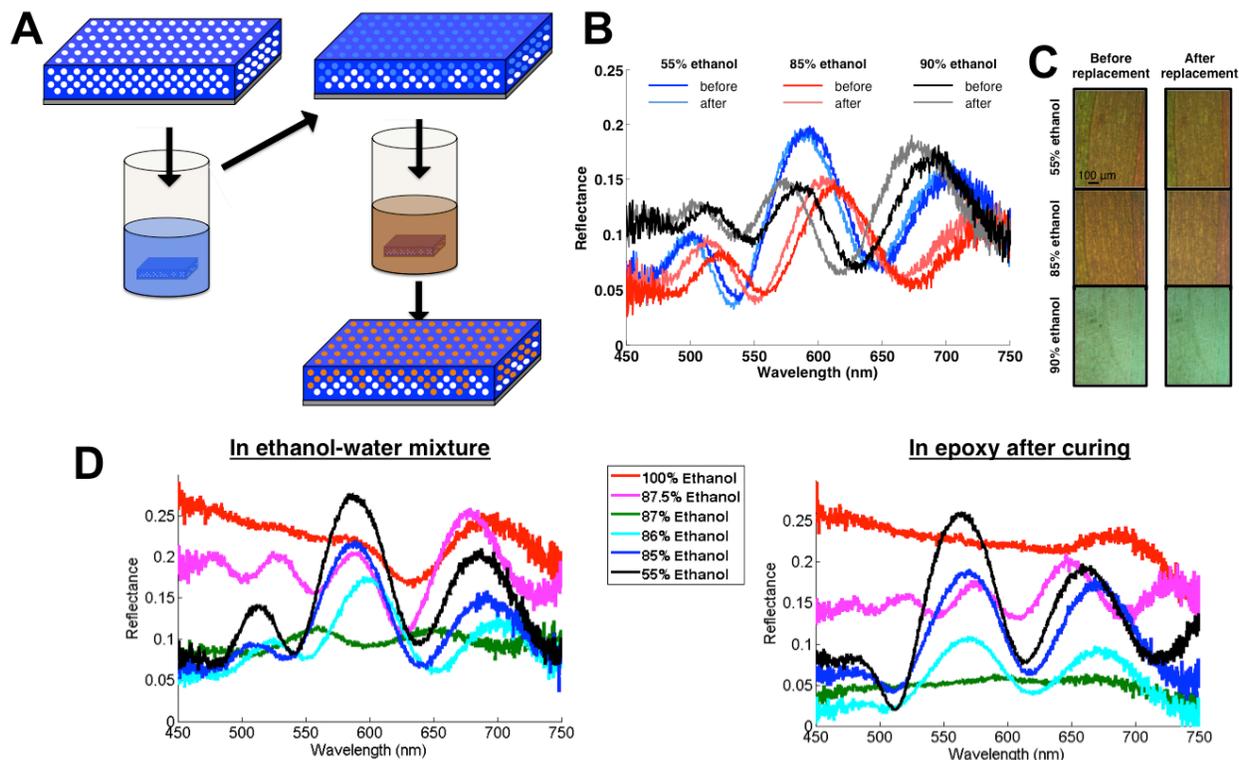

**Figure 5 – Freezing pore-filling profile at arbitrary degrees of partial wetting.** (A) Schematic showing the procedure. An IOF is immersed in an ethanol-water mixture whose ethanol content is fine-tuned to produce the desired state of partial wetting. Without drying the IOF, the mixture is exchanged with OG 142 epoxy resin. The IOF is then removed from the epoxy resin and the resin is UV cured. (B,C) This freezing procedure relies on the free-energy dynamics of wetting in IOF pores *(32)*, which do not favor pore de-wetting as the liquid surface tension increases. Spectra (B) and optical images (C) of three partial wetting patterns in a DEC-functionalized IOF (9 layers) induced by immersing the dry structure into three different ethanol-water mixtures ("before") and maintained after the wet IOFs are immersed in excess water ("after"). (D) Spectra of IOFs (9-layers) taken while immersed in different mixtures of ethanol and water (left) and then after the ethanol-water mixture is exchanged for epoxy resin and then cured (right).

We developed an analogous approach to permanently freeze in place disordered photonic structures produced by partial wetting. This process is shown schematically in Fig. 5A. This method exploited the fact that wetting in IOFs is limited by metastable pinning at the necks and not a global free-energy minimum *(31,32)*. As a result, we can exchange a liquid that has partially wet an IOF with a second miscible liquid without changing the partial wetting profile as long as the intrinsic contact angle does not decrease during the mixing process *(32)*. Figure 5B,C shows a proof of this concept. We wetted an IOF (9 layers in total thickness) in various mixtures of ethanol in water that produced different degrees of wetting (55%, 85% and 90% ethanol) and then purged them in water without drying first. This resulted in no significant changes to either the reflection spectrum or the color of the films (aside from very small shifts due to the slight refractive-index difference between the ethanol-water mixtures and pure water).

We exploited this effect to freeze partial wetting patterns produced in different ethanol-water mixtures, by exchanging the mixture with OG 142 epoxy resin (EPO-TEK) instead of water. Like water, this epoxy resin was found to not infiltrate the pores when exposed to dry DEC-functionalized IOFs, indicating a wettability to the pore surfaces that was always poorer than the starting ethanol-water mixture, the critical requirement for maintaining the filling state dictated by the first mixture. To ensure minimal disturbance of the wetting state and consistent miscibility with the epoxy during the exchange, we first exchanged the starting ethanol-water mixture (55-100% ethanol) with a 50:50 ethanol-water mixture and then performed a second exchange with the epoxy resin. After allowing sufficient time for liquid exchange to occur, we UV-cured the resin, solidifying the partially wet IOF. Fig. 5D,E show the reflection spectra of segments of an IOF (9 layers in total thickness) in ethanol-water mixtures that produce various different partial filling patterns (Fig. 5D) and the same films after these mixtures have been replaced with epoxy and cured (Fig. 5E), illustrating that the full range of partial wetting patterns can be frozen in epoxy. Fig. S12 compares the spectra for each sample in the ethanol-water mixtures and in the epoxy resin before and after curing. In contrast to the ethanol-water to water exchange (Fig. 5B), there is a considerable refractive difference between the epoxy ($n$ = 1.58) and the ethanol-water mixture ($n \sim 1.35$) that causes the spectral features to significantly blueshift upon exchange. No shifts of spectral features were observed during curing, indicating that the epoxy remains effectively pinned in place and expected volume changes during curing are accommodated by excess epoxy resin present above the structure.

The methods we have introduced here to dynamically control and freeze into place disorder in 3D photonic crystals stand out for their simplicity, given the versatility of photonic properties they produce. In low-refractive-index systems such as the one studied here, this technique may be used to tune the appearance of structural colors (brightness, hue and diffusivity) in a way that is compatible with large-scale manufacturing. We expect that applying this technique to 3D photonic crystals with higher refractive index contrast, sufficient to produce complete 3D photonic band gaps, may lead to the observation of several additional interesting phenomena such as spontaneous cavity formation and random lasing. These results illustrate the versatility of interfacial phenomena in directing and tuning self-assembly of hierarchical and aperiodic structures, mirroring the bottom-up symmetry breaking strategies employed by biological systems in morphogenesis.

**Acknowledgements**
The authors thank Mackenzie Kinney, Kevin Raymond, Natalie Koay and Prof. Mathias Kolle for helpful discussions. This work was supported by the US Air Force Office of Scientific Research Multidisciplinary University Research Initiative under Award FA9550-09-1-0669-DOD35CAP. IBB acknowledges support from a Banting Postdoctoral Fellowship funded by the Natural Sciences and Engineering Research Council of Canada.

**References**
1. J. D. Joannopoulos, S. G. Johnson, J. N. Winn, R. D. Meade, *Photonic crystals, molding the flow of light* (Princeton Univ. Press, Princeton, NJ, ed. 2, 2011).
2. J. D. Joannopoulos, P. R. Villeneuve, S. Fan, Photonic crystals: putting a new twist on light. *Nature* **386,** 143-149 (1997).
3. S. John, Strong localization of photons in certain disordered dielectric superlattices. *Phys. Rev. Lett.* **58,** 2486-2489 (1987).
4. E. Yablonovitch, Inhibited spontaneous emission in solid-state physics and electronics. *Phys. Rev. Lett.* **58,** 2059-2062 (1987).
5. M. Campbell, D. N. Sharp, M. T. Harrison, R. G. Denning, A. J. Turberfield, Fabrication of photonic crystals for the visible spectrum by holographic lithography. *Nature* **404,** 53-56 (2000).
6. S. Noda, K. Tomoda, N. Yamamoto, A. Chutinan, Full three-dimensional photonic bandgap crystals at near-infrared wavelengths. *Science* **289,** 604-606 (2000).
7. B. H. Cumpston *et al.*, Two-photon polymerization initiators for three-dimensional optical data storage and microfabrication. *Nature* **398,** 51-54 (1999).
8. G. von Freymann, V. Kitaev, B. V. Lotsch, G. A. Ozin, Bottom-up assembly of photonic crystals. *Chem. Soc. Rev.* **42,** 2528-2554 (2013).
9. C. I. Aguirre, E. Reguera, A. Stein, Tunable colors in opals and inverse opal photonic crystals. *Adv. Func. Mater.* **20,** 2565-2578 (2010).
10. Y. A. Vlasov, X.-Z. Bo, J. C. Sturm, D. J. Norris, On-chip natural assembly of silicon photonic bandgap crystals. *Nature* **414,** 289293 (2001).
11. A. Blanco *et al.*, Large-scale synthesis of a silicon photonic crystal with a complete three-dimensional bandgap near 1.5 micrometres. *Nature* **405,** 437-440 (2000).
12. K. Ishizaki, M. Koumura, K. Suzuki, K. Gondaira, S. Noda, Realization of three-dimensional guiding of photons in photonic crystals. *Nat. Photonics* **7,** 133-137 (2013).
13. S. A. Rinne, F. García-Santamaría, P. V. Braun, Embedded cavities and waveguides in three-dimensional silicon photonic crystals. *Nat. Photonics* **2,** 52-56 (2008).

14. P. V. Braun, S. A. Rinne, F. García-Santamaría, Introducing defects in 3D photonic crystals: state of the art. *Adv. Mater.* **18,** 2665-2678 (2006).
15. D. S. Wiersma, P. Bartolini, A. Lagendijk, R. Righini, Localization of light in a disordered medium. *Nature* **390,** 671-673 (1997).
16. T. Schwartz, G. Bartal, S. Fishman, M. Segev, Transport and Anderson localization in disordered two-dimensional photonic lattices. *Nature* **446,** 52-55 (2007).
17. Y. Lahini *et al.*, Anderson localization and nonlinearity in one-dimensional disordered photonic lattices. *Phys. Rev. Lett.* **100,** 013906 (2008).
18. P. D. García, R. Sapienza, C. Toninelli, C. López, D. S. Wiersma, Photonic crystals with controlled disorder. *Phys. Rev. A* **84**, 023813 (2011).
19. L. Sapienza *et al.*, Cavity quantum electrodynamics with Anderson-localized modes. *Science* **327,** 1352-1355 (2010).
20. J. Liu *et al.*, Random nanolasing in the Anderson localized regime. *Nature Nanotechnol.* **9,** 285-289 (2014).
21. C. Rockstuhl, T. Scharf, Eds. *Amorphous nanophotonics* (Springer-Verlag, Berlin 2013).
22. E. R. Martins *et al.*, Deterministic quasi-random nanostructures for photon control. *Nat. Commun.* **4,** 2665 (2013).
23. Y. S. Chan, C. T. Chan, Z. Y. Liu, Photonic band gaps in two dimensional photonic quasicrystals. *Phys. Rev. Lett.* **80,** 956–959 (1998).
24. A. Ledermann *et al.*, Three-dimensional silicon inverse photonic quasicrystals for infrared wavelengths. *Nat. Mater.* **5,** 942–945 (2006).
25. P. Vukusic, J. R. Sambles, Photonic structures in biology. *Nature* **424,** 852-855 (2003).
26. P. Vukusic, B. Hallam, J. Noyes, Brilliant whiteness in ultrathin beetle scales. *Science* **315,** 348 (2007).
27. C. Pouya, D. G. Stavenga, P. Vukusic, Discovery of ordered and quasi-ordered photonic crystal structures in the scales of the beetle Eupholus magnificus. *Opt. Express* **19,** 11355-11364 (2011).
28. R. O. Prum, R. H. Torres, S. Williamson, J. Dyck, Coherent light scattering by blue feather barbs. *Nature* **396,** 28-29 (1998).
29. R. O. Prum, J. A. Cole, R. H. Torres, Blue integumentary structural colours in dragonflies (Odonata) are not produced by incoherent scattering. *J. Exp. Biol.* **207,** 3999–4009 (2004).
30. V. Saranathan *et al.*, Structural diversity of arthropod biophotonic nanostructures spans amphiphilic phase-space. *Nano Lett.* DOI: 10.1021/acs.nanolett.5b00201 (2015).
31. I. B. Burgess *et al.*, Encoding complex wettability patterns in chemically functionalized 3D photonic crystals. *J. Am. Chem. Soc.* **133,** 12430-12432 (2011).
32. I. B. Burgess *et al.*, Wetting in color: colorimetric differentiation of organic liquids with high selectivity. *ACS Nano* **6,** 1427-1437 (2012).
33. B. Hatton, L. Mishchenko, S. Davis, K. H. Sandhage, J. Aizenberg, Assembly of large-area, highly ordered, crack-free inverse opal films. *PNAS* **107,** 10354-10359 (2010).
34. A. Tuteja, W. Choi, J. M. Mabry, G. H. McKinley, R. E. Cohen, Robust omniphobic surfaces. *PNAS* **105,** 18200-18205 (2008).
35. D. Wilkinson, J. F. Willemsen, Invasion percolation: a new form of percolation theory. *J. Phys. A Math. Gen.* **16,** 3365-3376 (1983).
36. M. Prat, Isothermal drying on non-hygroscopic capillary-porous materials as an invasion percolation process. *Int. J. Multiphase Flow* **21,** 875–892 (1995).
37. T.-S. Wong *et al.*, Bioinspired self-repairing slippery surfaces with pressure-stable omniphobicity. *Nature* **477,** 443-447 (2011).
38. J. D. Smith *et al.*, Droplet mobility on lubricant-impregnated surfaces. *Soft Matter* **9,** 1772-1780 (2013).
39. G. England *et al.,* Bioinspired micrograting arrays mimicking the reverse color diffraction elements evolved by the butterfly Pierella luna, *PNAS* **111**, 15630-15634 (2014).

## Methods

**IOF synthesis**
Inverse-opal films (IOFs) were deposited on silicon substrates using a colloidal co-assembly technique described previously *(31-33)*. After fabrication, the surfaces of all IOFs were functionalized with n-decyl trichlorosilane (DEC) as described previously *(32)*.

**Drying experiments**
Each drying experiment started with immersing a DEC-functionalized IOF in an aliphatic liquid (decane, undecane, dodecane and decane-PDMS mixtures were used in different experiments, as indicated in the main text) leading to complete wetting of all pores. Once pores had completely filled (as observed via the color *(32)*), the IOF was removed and immediately placed under running DI water for 5-10 seconds. The running DI water, immiscible with the aliphatic liquid and having a poorer affinity for the DEC-functionalized pore surfaces than it, served to clean the surface of any excess aliphatic liquid. The trapped aliphatic liquid that remained trapped in the pores formed a lubricating thin film that allowed all excess water to slide off *(37,38)*. The IOF was then placed under a modified optical microscope (Leica) that had the capacity to take bright-field (BF), dark-field (DF) and Bertrand-lens (BL) images as well as reflectance spectra (RS) *(39)*. Subsequent drying experiments were done on the same region of the IOF to capture the full time-evolution in BF, DF, BL, and RS. Dodecane was chosen for the bulk of the characterization experiments because of its volatility, which was low enough to give ample time after DI water flushing had completed to set up the imaging experiment before the pores had begun to empty.

Before creating a frozen pattern on an IOF, a complete dodecane drying experiment was run with the time evolution of the RS recorded to use as the comparison reference (Fig. 4B). PDMS (Sylgard 184, 10:1 base:cross-linker) was prepared and mixed with decane at different ratios. After completing the dodecane experiment, the IOF was submerged in a chosen PDMS-decane solution and then placed under running DI water as in normal drying experiments. The sample was then left at room temperature for 3-5 hours, ensuring the volatile component had evaporated, and then placed in an oven at 70°C overnight to allow the PDMS to cure. Samples were characterized after PDMS curing.

**Partial wetting experiments**
Partial wetting in DEC-functionalized IOFs was characterized (BF, DF, BL, RS) while they were submerged in a small dish containing mixtures of water and ethanol. The solvent exchange tests (Fig. 5B,C) were carried out by first imaging the IOF in the starting ethanol-water mixture and then quickly removing the IOF and submerged in a second dish containing DI water. To freeze partial wetting patterns, we submerged the IOF first in the desired ethanol-water mixture. We then rapidly removed it and submerged it in a 50% ethanol solution to prime it for epoxy exchange. After leaving it to rest for at least 5 minutes to ensure equilibration in the pores, we removed the IOF from this solution and rapidly submerged it (before drying could occur) in a container containing excess UV-curable epoxy resin (EPO-TEK OG 142). After ~30 s of immersion in the resin solution with gentle agitation, the IOF was removed from the resin and sandwiched between two thin layers of cured PDMS. This is done to squeeze out most of the excess epoxy on top of the IOF surface and provide more consistent imaging conditions before and after curing. After imaging and spectral characterization of the uncured samples they were then cured under a UV lamp for 30 minutes.

**Simulations**
Simulations were built based on the model described previously for partial wetting in inverse opals *(32)*. To facilitate a computationally efficient simulation over as large an area as possible (large areas were needed to adequately capture the statistics of percolation phenomena) with results that were easy to visualize, a 2D IOF lattice (hexagonal) was used. The model generated a lattice with a length of 200 units and a thickness of 25 layers. Nearest neighbors were connected by necks. Neck angles ($\phi_0$) were randomly assigned to each nearest-neighbor connection according to a normal distribution with a mean and standard deviation of 19.6° and 3.2° respectively, matching typical experimental values previously measured *(32)*. To calculate a partial wetting pattern, we first assigned an intrinsic contact angle ($\theta_c$) for a liquid that was close to the mean neck angle (see Fig. 1A for values used). Then, starting with a completely filled top layer (as these are half-spheres in our IOFs and have no re-entrant curvature to pin the meniscus *(31,32)*), the simulation filled all pores with paths of interpore fluid connectivity (i.e., where $\phi_0 > \theta_c$) connecting them to the top filled layer. Drying patterns were simulated from an initial condition of a completely empty top layer (half-pores) with all other pores filled. The simulation then emptied the pores one-by-one, emptying at each step the filled pore associated with the largest neck angle (weakest pinning effect) that connected a filled and empty pore in the current configuration.

For each filling state analyzed (partial wetting or drying), a refractive index map of the structure was created with a spatial resolution of 1/30 of the lattice spacing using a liquid refractive index of 1.35 and a matrix refractive index of 1.4. Fourier transforms (FTs) of these maps were also taken, where the image was first made square by adding a substrate with the matrix refractive index of sufficient thickness below. These maps are shown in Fig. 1A,B. Filling vs. depth maps (Fig. 1C,D) and disorder profiles as a function of total filling fraction (Fig. 1E,F) reflect the average result of 10 successive simulations, where a new random neck distribution is generated for each. This was done to eliminate artifacts arising from particular neck distributions. The disorder parameter (Fig. 1E,F) was calculated from the FTs using the fraction of the total spectral density that lay within the first Brillouin zone, but outside of the vertical axis.

## Supplementary Figures

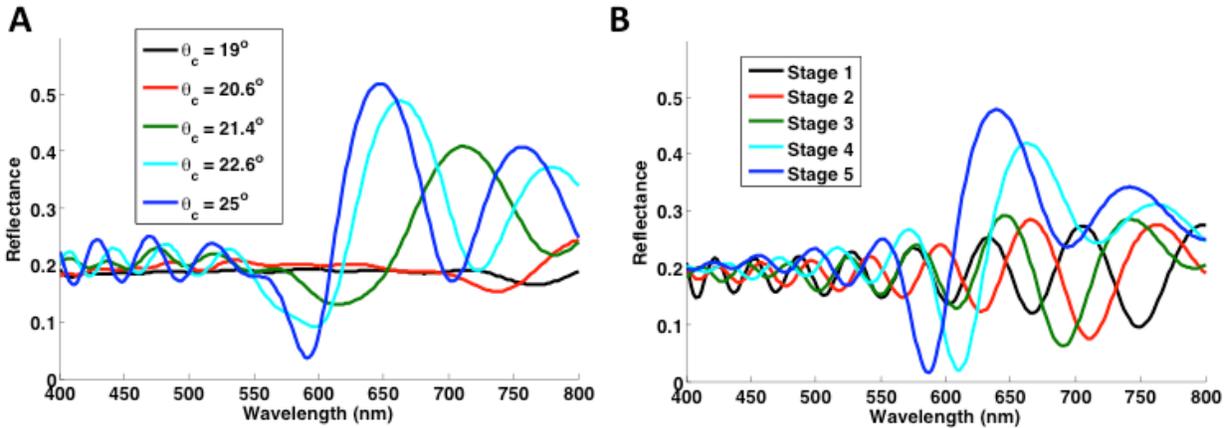

**Fig. S1 – 1D simulations of the spectral percolation response.** Simulated reflectance calculated using the transfer matrix method for 1D refractive-index profiles generated by 2D percolation simulations (Fig. 1A,B, main text) for partial wetting (A) and drying (B). 1D refractive-index profiles were generated by averaging the dielectric constant 2D refractive-index profiles across the horizontal dimension. Redshifting of spectral features occurs with increased filling in both sets of simulations, as is observed experimentally. However, the 1D simulations do not capture the primary effect of transverse disorder: the dramatic drop and recovery in the total reflectance that occurs as percolation progresses.

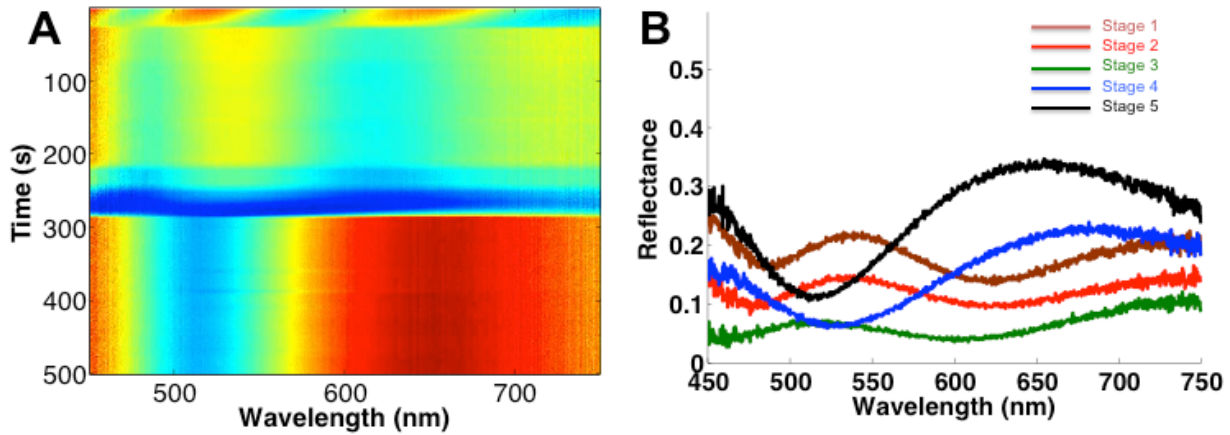

**Fig. S2 – Time evolution of reflectance during drying (3-layer film).** (A) Time evolution of reflectance at normal incidence as dodecane dries. (B) Spectra at the five stages defined by total reflectance (see main text).

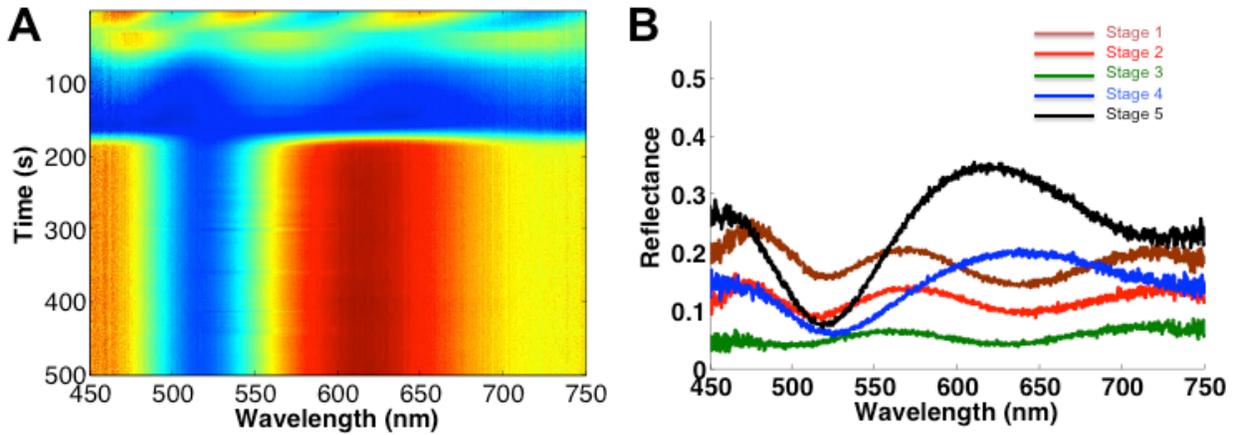

**Fig. S3 - Time evolution of reflectance during drying (4-layer film).** (A) Time evolution of reflectance at normal incidence as dodecane dries. (B) Spectra at the five stages defined by total reflectance (see main text).

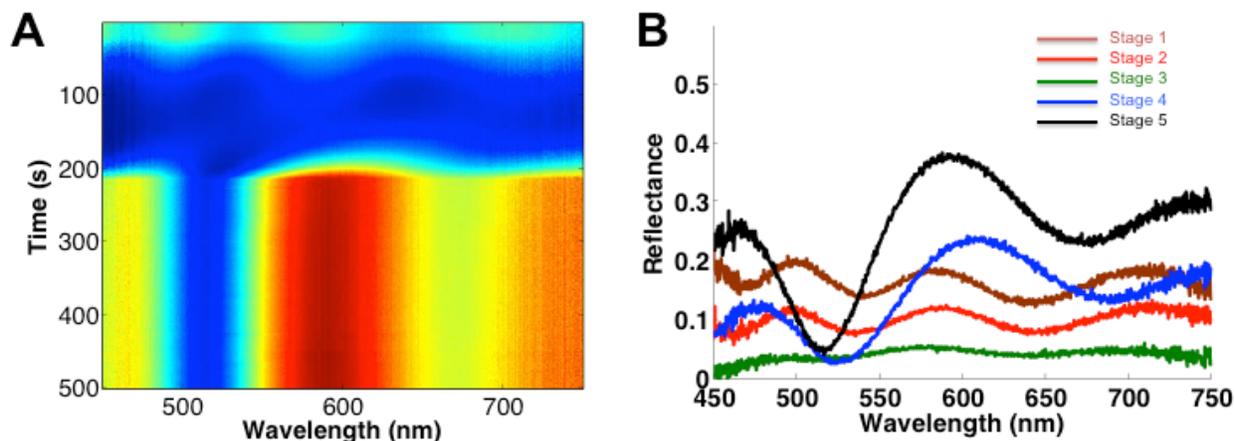

**Fig. S4 - Time evolution of reflectance during drying (5-layer film).** (A) Time evolution of reflectance at normal incidence as dodecane dries. (B) Spectra at the five stages defined by total reflectance (see main text).

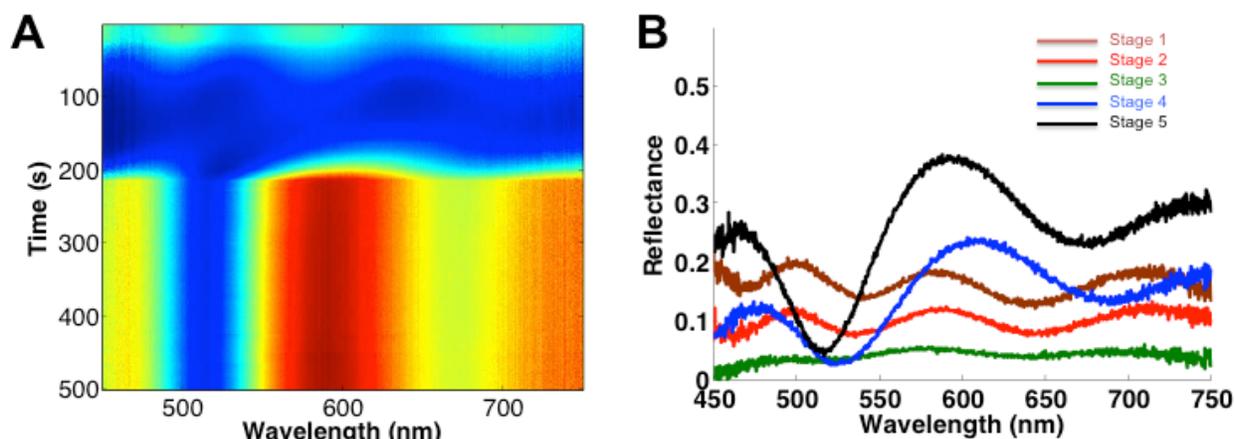

**Fig. S5 - Time evolution of reflectance during drying (6-layer film).** (A) Time evolution of reflectance at normal incidence as dodecane dries. (B) Spectra at the five stages defined by total reflectance (see main text).

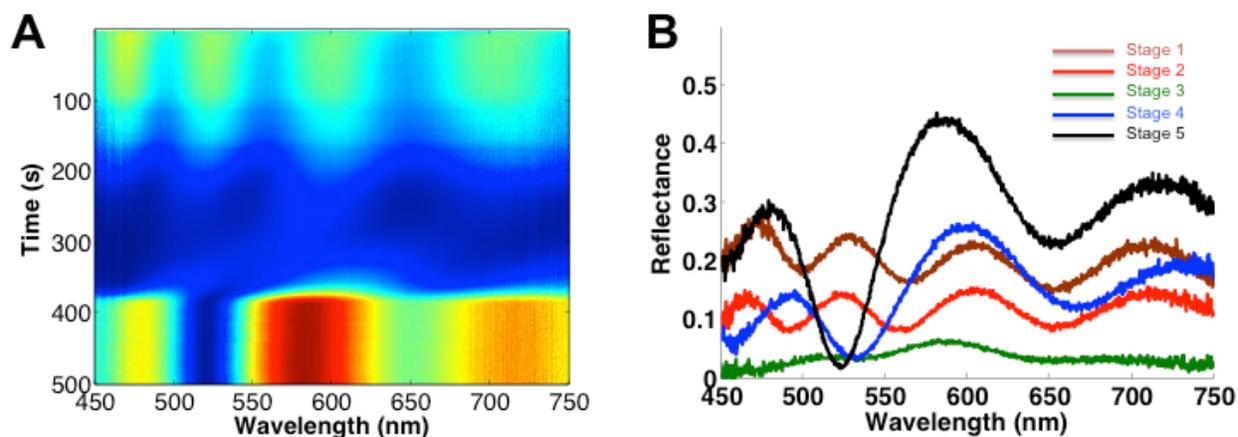

**Fig. S6 - Time evolution of reflectance during drying (7-layer film).** (A) Time evolution of reflectance at normal incidence as dodecane dries. (B) Spectra at the five stages defined by total reflectance (see main text).

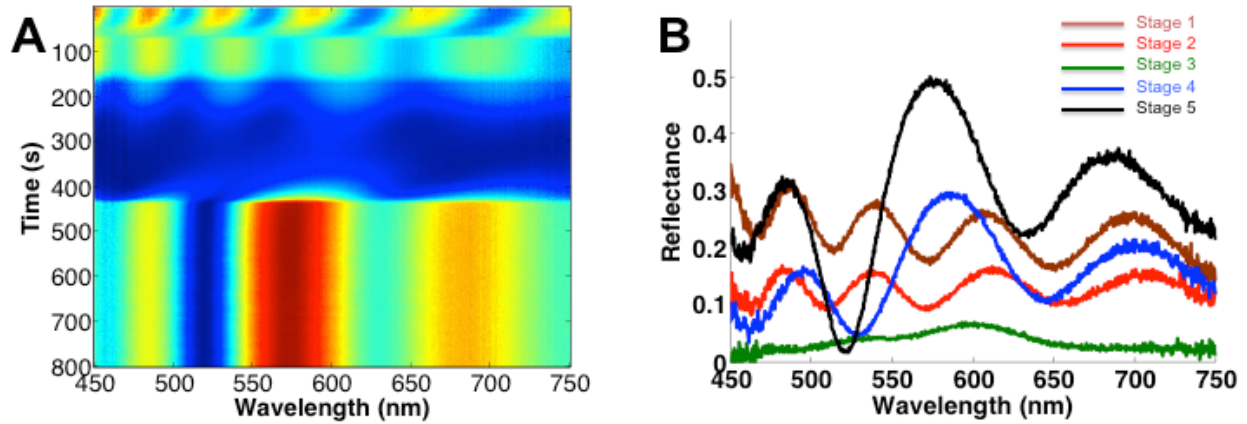

**Fig. S7 - Time evolution of reflectance during drying (8-layer film).** (A) Time evolution of reflectance at normal incidence as dodecane dries. (B) Spectra at the five stages defined by total reflectance (see main text).

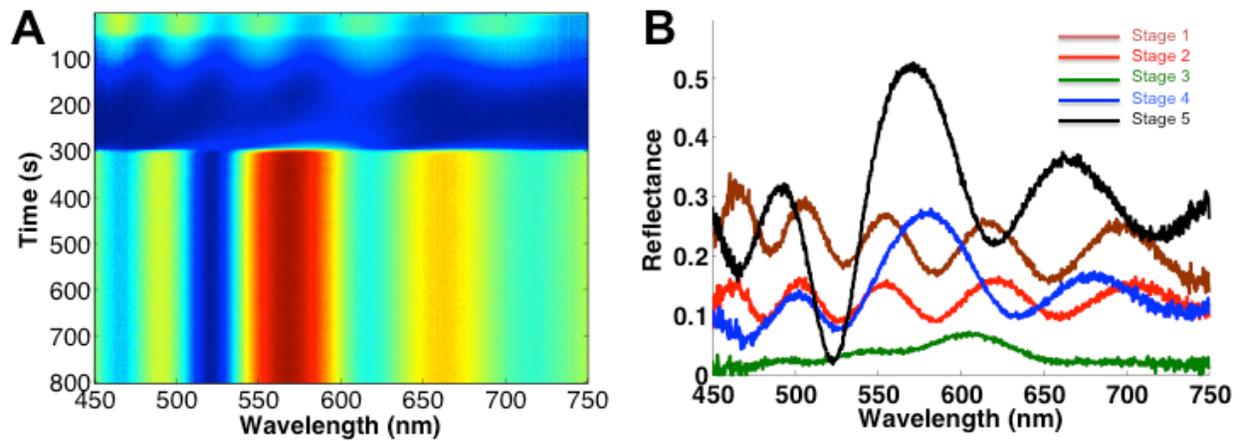

**Fig. S8 - Time evolution of reflectance during drying (9-layer film).** (A) Time evolution of reflectance at normal incidence as dodecane dries. (B) Spectra at the five stages defined by total reflectance (see main text).

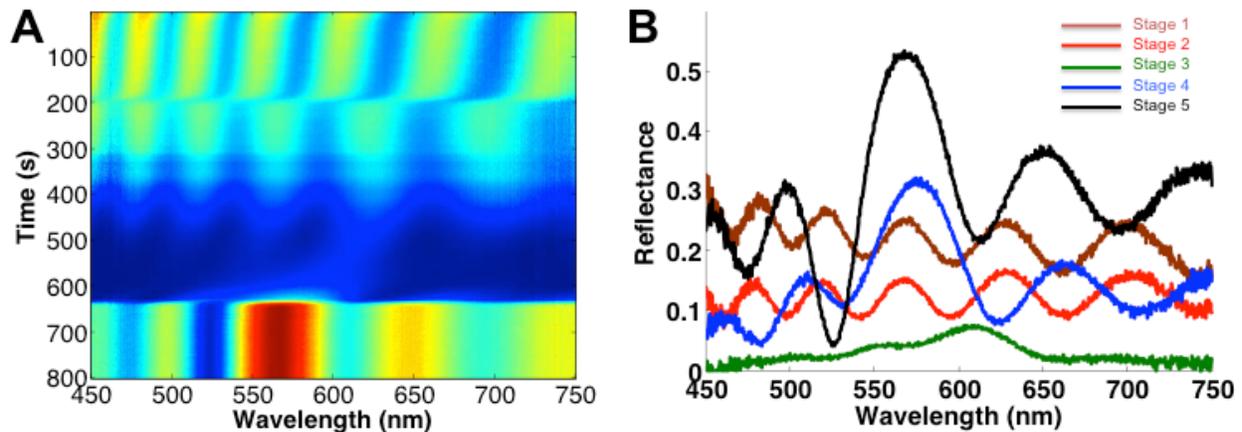

**Fig. S9 - Time evolution of reflectance during drying (10-layer film).** (A) Time evolution of reflectance at normal incidence as dodecane dries. (B) Spectra at the five stages defined by total reflectance (see main text).

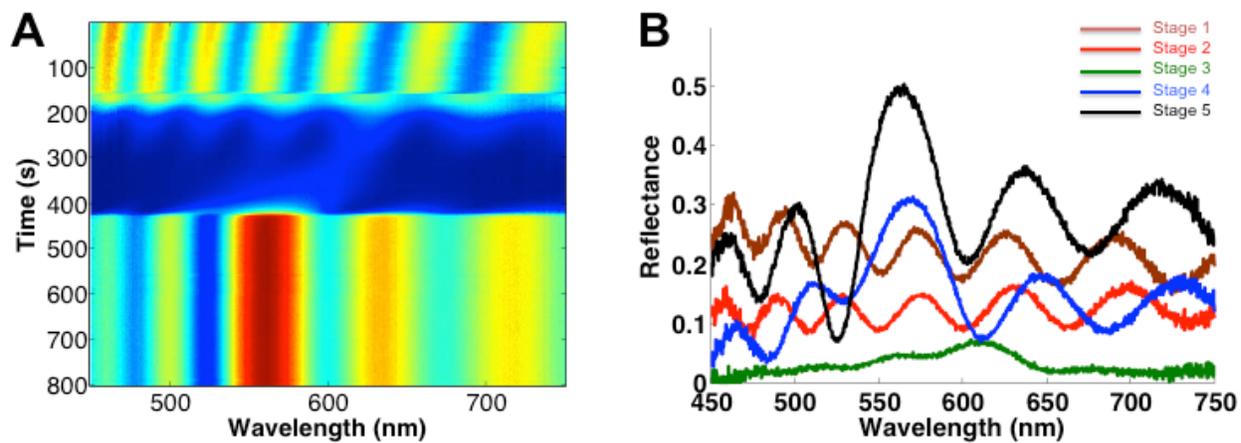

**Fig. S10 - Time evolution of reflectance during drying (11-layer film).** (A) Time evolution of reflectance at normal incidence as dodecane dries. (B) Spectra at the five stages defined by total reflectance (see main text).

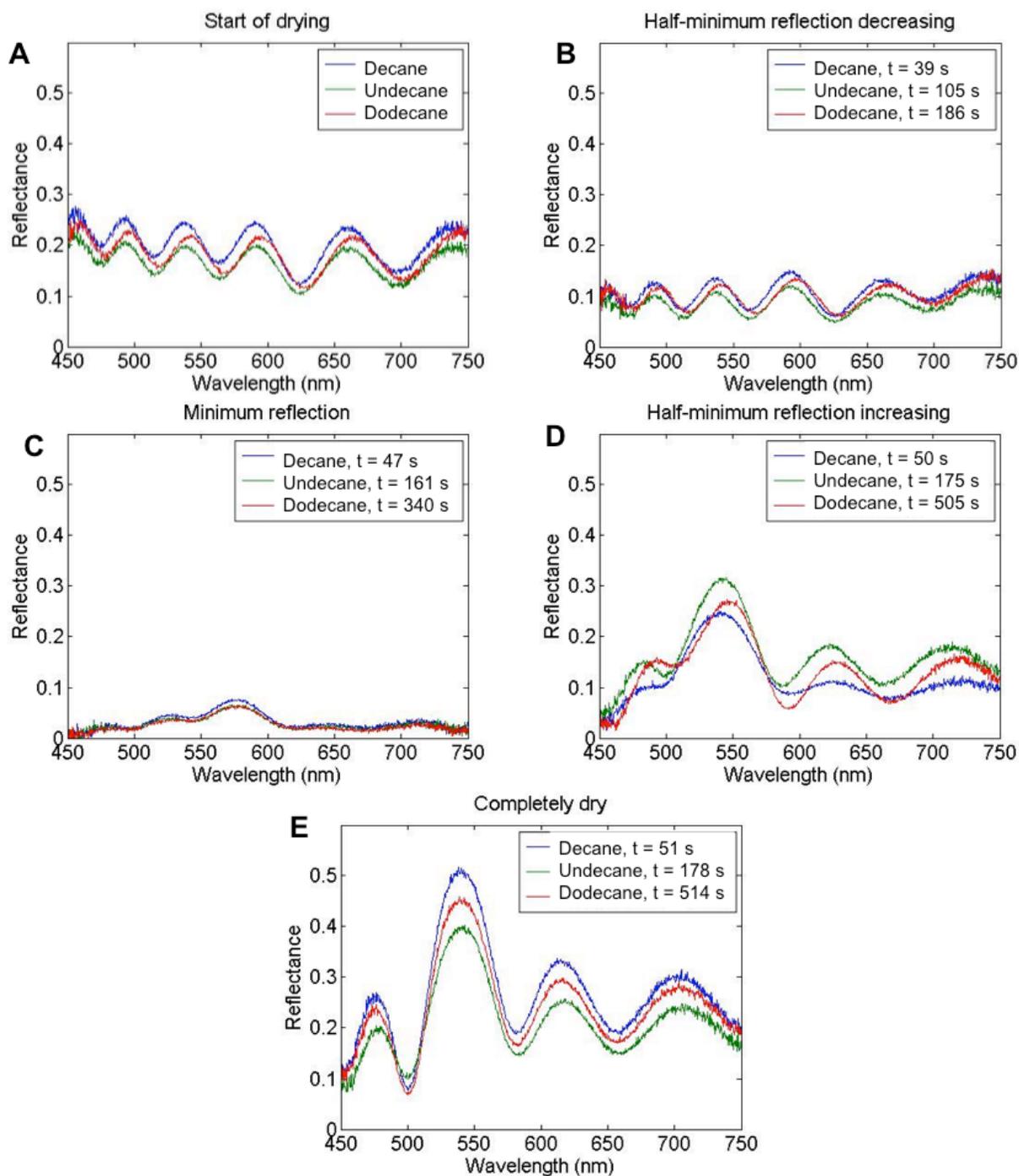

**Figure S11 – Drying sequence in different liquids.** Comparison of normal-incidence reflection spectra of an IOF (9 layers) at the five stages of drying: (A) Stage 1: start of drying; (B) Stage 2: total reflectance decreases to half of the minimum with respect to stage 1; (C) Stage 3: point of minimum total reflectance; (D) Total reflectance recovered to halfway between the minimum and that of a dry film; Stage 5: dry film.

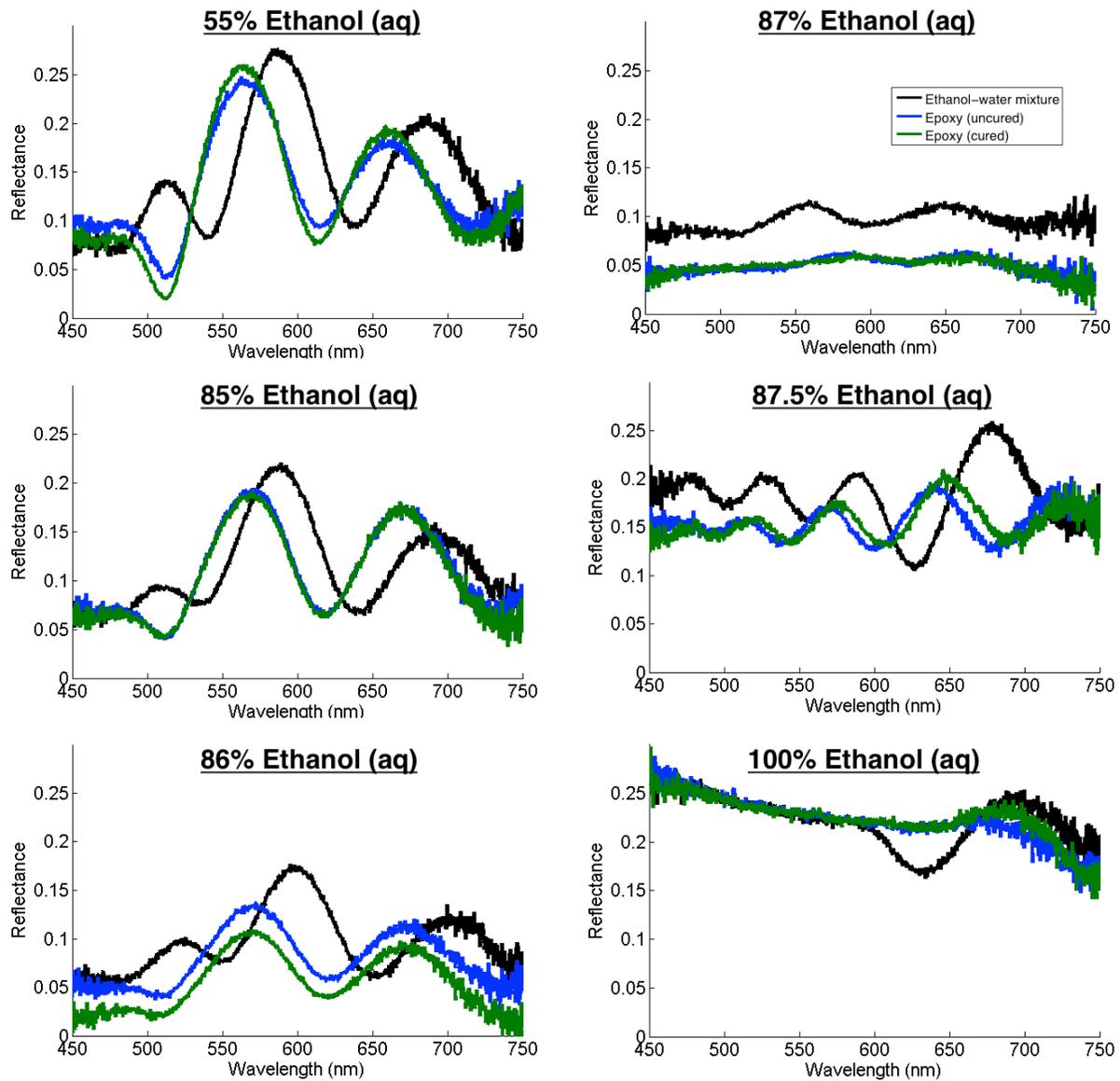

**Figure S12 – Freezing of partial wetting via epoxy resin exchange.** Reflectance spectra at normal incidence taken of IOFs (9-layers) after immersion in an ethanol water-mixture (black curves), after exchange of the mixture with epoxy resin (blue curves) and then after the epoxy resin (OG 142) has been cured (green curves). The refractive-index difference between the epoxy (n = 1.58) and the ethanol-water mixture (n = 1.35) causes the spectral features to blueshift.